\documentclass[12pt]{iopart}
\usepackage[pdftex]{graphicx}
\newcommand{\cm}{cm$^{-1}$}
\begin{document}
\title[Micro-Raman and FE studies of SINWs...]
{Micro-Raman and field emission studies of silicon nanowires prepared by metal assisted chemical etching}
\author{Vivek Kumar$^1$ \footnote{Electronic mail: vivek@unitus.it}, \ Shailendra K. Saxena$^2$, Vishakha Kaushik$^3$, Kapil Saxena$^3$, Rajesh\  Kumar$^{2,4}$\  and A.K. Shukla$^3$}

\address{$^1$ Biophysics \& Nanoscience Centre, DEB-CNISM, Universita della Tuscia, I-01100 Viterbo, Italy}

\address{$^2$ Discipline \  of \  Physics, \ School\  of \  Basic \ Sciences, \ Indian \ Institute \  of \ Technology \ Indore, \ Madhya \ Pradesh-452017, \ India}
\address{$^3$ Department \  of \  Physics,  \ Indian \ Institute \  of \ Technology \ Delhi, \ Hauz \ Khas-110016, \ India}
\address{$^4$ Material \  Science  \ and \ Engineering \  Group, \ Indian \ Institute \  of \ Technology \ Indore, \ Madhya \ Pradesh-452017, \ India }

\begin{abstract}

Micro-Raman scattering and electron field emission characteristics of silicon nanowires (SiNWs) synthesized by metal assisted chemical etching (MACE) are investigated. Scanning electron microscopy images reveal the growth of well aligned vertical SiNWs. Raman shift and size relation from bond-polarizability model has been used to calculate exact confinement sizes in SiNWs. The Si optical phonon peak for SiNWs showed a downshift and an asymmetric broadening with decreasing diameter of the SiNWs due to quantum conﬁnement of optical phonons. The field emission characteristics of these SiNWs are studied based by carrying out current–voltage measurements followed by a theoretical analysis using Fowler–Nordheim equation. The electron field emission increased with decreasing diameter of SiNWs. Field emission from these SiNWs exhibits signiﬁcant enhancement in turn-on ﬁeld and total emission current with decreasing nanowire size. The reported results in the current study indicate that MACE is a simple technique to prepare well-aligned SiNWs with potentials for applications in field emission devices.

\end{abstract}

\maketitle

\section{Introduction}
Silicon (Si) nanowires (NWs) have attracted great interest due to their compatibility with silicon-based electronic technology and their various potential applications in many fields, for instance nanoelectronics [1], optoelectronics [2] energy conversion [3,4] and energy storage [5] as well as bio and chemical sensors [6]. One of the reasons for the versatile application of nanomaterials is ``quantum confinement effect” which offers a great flexibility of tailoring optical, electrical, thermal, and chemical properties of these materials for desired purposes. These one-dimensional systems are also ideal systems for exploring a large number of novel phenomena at the nanoscale and investigating the size and dimensionality dependence of their properties for fundamental research [7–9] This kind of size dependent change in various properties of nanomaterials should be investigated in details for extending the application of nanomaterials in new fields. In addition, investigating various properties of nanomaterials will enhance the basic mechanism behind a particular observation exhibited by the nanomaterial.

Phonon Raman spectra of semiconductor nanostructures became a standard and very powerful tool to calculate the size of nanostructures [10–12] In a crystalline material, the region over which the phonon wave function extends is infinite and only zone centre phonons (q$\sim$0) contribute to first-order Raman scattering. In nanostructures, phonons are confined to a small size (of the order of few nanometers). Due to this, translational symmetry of the ideal crystal is disturbed and the conservation law of momentum is relaxed. Therefore, the optical phonons, away from the zone center (q$\geq$0) can also contribute to the Raman scattering process [13]. This results in the softening and asymmetric broadening of the optical phonon Raman modes. Several models have been developed to quantify the crystal size from such Raman-shift and Raman line-shape. The most commonly used are the bond-polarizability model (BPM)[14], where a simple relation between SiNWs size and peak shift is derived, and the phenomenological quantum confinement model [10,15] that uses an ad-hoc confinement function, allows to calculate the line shape of the transverse optic one-phonon mode as a function of the SiNWs size. In the present manuscript, BPM has been used to estimate size of SiNWs under investigation as will be discussed later.

Similar to carbon nanotubes, a very good field emitter, it is of interest to investigate the field emission properties of SiNWs. Many efforts have been made to investigate the electron field emission properties of SiNWs, which have promising applications in field emission displays, cold cathode electron sources and microwave devices [16,17]. Field electron emission properties of taperlike and rodlike SiNWs [18,19], H-passivated SiNWs [20], sponge-like SiNW [21], Si nanotubes [22] and other SiNWs [23–-26] have been reported. These NWs were prepared by different methods such as vapor–liquid–solid process, chemical vapor deposition, thermal evaporation, laser ablation etc. however, their growth mechanisms have some limitations including high temperature or vacuum conditions, special templates and complex equipments[19, 27–-29]. Recently, metal assisted chemical etching (MACE) method was developed to prepare lager-area aligned SiNWs arrays on silicon substrates close to room temperature[30–-32]. MACE is a simple, low-expense and better growth orientation method of fabricating large scale SiNW arrays. To date, the electron field emission characteristics of these vertically aligned SiNWs fabricated by MACE have been rarely investigated and thus open scope for further studies, having potential applications in field emission devices.

Main aim of the present paper is to investigate the effect of quantum confinement effect on field emission properties of SiNWs by studying size dependent field emission characteristics have been studied in these SiNWs. The SiNWs under investigation have been synthesized on silicon substrates by metal assisted chemical etching technique. Optical phonon Raman spectra are used here to calculate the mean sizes of SiNWs using bond-polarizability model which suggest presence of quantum confinement effect. A correlation between SiNWs’ size and its field emission properties show that turn on voltage also is affected by quantum confinement effect in SiNWs.

\section{Experimental details}
Si NWs samples were prepared by metal assisted chemical etching (MACE) of n-Si (100) wafer having resistivity of 3-5 $\Omega$-cm. These wafers were cleaned in acetone and ethanol to remove impurities prior to starting the MACE process. The cleaned wafers were immersed in HF solution to remove thin oxide layer formed at surface. These wafers were dipped in solution containing 4.8 M HF \& 5 mM AgNO$_3$ for one minute at room temperature to deposit Ag nano particles (Ag NPs). The Ag NPs deposited samples were then kept for etching in an etching solution containing 4.6 M HF \& 0.5 M H$_2$O$_2$. Etched wafers were transferred in HNO$_3$ acid to dissolve Ag metal present on the etched sample. Then the samples were dipped into HF solution to remove oxide layer induced by HNO$_3$ used in the previous step. Three samples S1, S2 and S3 were prepared for etching time 25 minutes, 35 minutes and 45 minutes respectively with all other parameters unchanged. Surface morphologies of all these samples have been characterized with a supra55 Zeiss Scanning Electron Microscope (SEM). Raman spectra were recorded using Jobin Horiba (T64000 triple monochromator) micro-Raman spectrometer in a backscattering geometry at room temperature. Raman scattering modes were excited by an argon ion laser ($\lambda$ = 514.5 nm). The ﬁeld emission (FE) properties were measured using a diode conﬁguration with the sample as a cathode and a stainless steel plate as an anode. The gap between the sample surface and the anode was about 250 $\mu$m. The FE measurements were performed in vacuum at a base pressure $\sim$ 10$^{-6}$ torr.

\section{Results and Discussion}

Figure 1 shows SEM images of SiNWs prepared by MACE for 45 min (sample S3). Figure 1(a) shows top view of sample S3, which indicates a uniform porous surface with distributed micro pores. The top-view suggests that a parallel well aligned pores of Si can be present. To understand the actual structure of the sample, a cross-sectional SEM imaging has been done as shown in fig 1(b). The cross-sectional image in Figure 1(b) shows that the sample contains aligned SiNWs type arrays which are uniform on the entire wafer surface and  can be seen that the nanowires are all straight, uniform and relatively vertical to the silicon substrate. Figure 1(c) shows the magnified cross-sectional SEM image of same sample, where a submicron sized nanowires appear to be formed. It is possible that the nanowires formed in thse samples are smaller than they appear in the SEM image and should be investigated using Raman spectroscopy to estimate the actual size of the nanostructure present in the sample. These results will be discussed later. The formation of SiNWs can be understood as follows. The Si atoms from the wafer get dissolved into HF acid in the presence of catalytic AgNPs.  The role of Ag is to catalyze the Si dissolution by both locally collecting electrons from Si and allowing Ag$^+$ ions to be reduced at its surface. The formation of Ag/Si Schottky junctions favours local etching near the Ag nanoparticles, hence the formation of vertical pores. The unetched walls between the pores constitute the nanowires as shown in Fig. 1. As clear from the above mechanism, the size of this undissolved wall can be controlled by etching time.

\begin{figure}[ht]\centering 
\includegraphics[width=8cm]{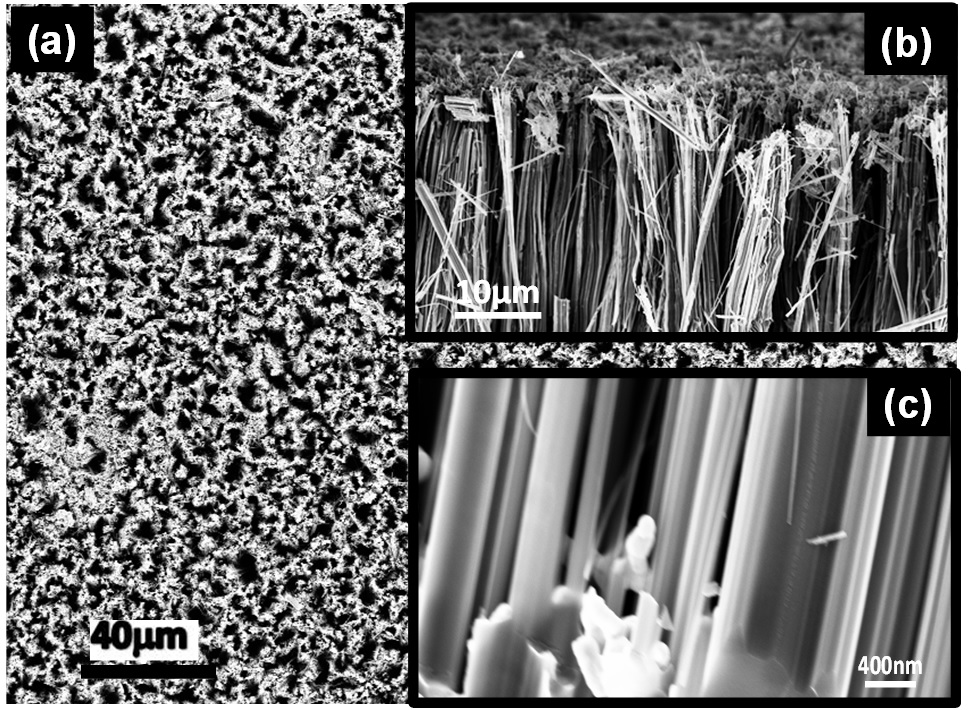}
\caption{SEM images of SiNWs synthesised by MACE for 45 minutes. (a) Plan view SEM image, (b) cross-sectional SEM image of the Si NW at low magnification showing the uniformity and NW length, (c) magnified cross-sectional SEM image of several NWs. }
\label{fig:view}
\end{figure}

Raman spectra from c-Si and samples S1, S2 \& S3 are shown in Figs. 2(a) – (d) respectively. Raman spectrum from the c-Si shows a symmetric Lorentzian Raman line-shape centered at 520 \cm with FWHM of 3.5 \cm as shown in Fig. 2(a) due to zone-center phonons present in the Si crystal. Raman spectra from samples S1 – S3 are shown in Figs. 2(b) - (d). Raman spectra of theses samples show following variations with increasing etching time. (1) Raman peak of the Si NSs is red-shifted as compared to usual Raman active optical mode of c-Si (i.e. 520 \cm). Raman peak shifts gradually from 520 \cm to 513.0 \cm as etching time is increased from 15 minutes to 45 minutes. (2) The Raman spectra are asymmetric in nature where asymmetry ratio increases with increasing etching time. Asymmetry ratio is defined here as $\frac{\Gamma_a}{\Gamma_b}$, where $\Gamma_a$ and $\Gamma_b$ are the half widths on the lower and higher sides of the Raman peak position. (3) Raman spectra get broadened in comparison to that of the c-Si.

The characteristics of the Raman spectra described above can be explained by the relaxation of momentum conservation due to quantum confinement of optical phonons [33]. Conservation of momentum in the c-Si allows only first-order Raman scattering of the optical phonons at the center of Brillouin-zone. It corresponds to a frequency shift of 520 \cm, as shown in Fig. 2(a). In SiNWs, the translational symmetry of the ideal crystal is disturbed and the conservation law of momentum is relaxed. If an optical phonon is confined within a space $\Delta L$, the momentum uncertainty $\Delta q$, is given by the relationship $\Delta L \Delta q \sim 1$. Therefore, the optical phonons near ($\Delta q \sim \frac{1}{\Delta L}$) the center of the Brillouin-zone become Raman active in quantum confined systems. Since the dispersion curve of optical phonons bends down towards the zone edge (q=1), the Raman peak shifts to lower frequency and broadens asymmetrically [13]. Thus, changes in the Raman line shape, compared to c-Si, is a function of the sizes of NWs. It is evident from the etching time dependent variation in Raman line-shape that samples S1-S3 contains SiNWs having size sufficiently low to show quantum confinement effect. 

\begin{figure}[ht]\centering 
\includegraphics[width=8cm]{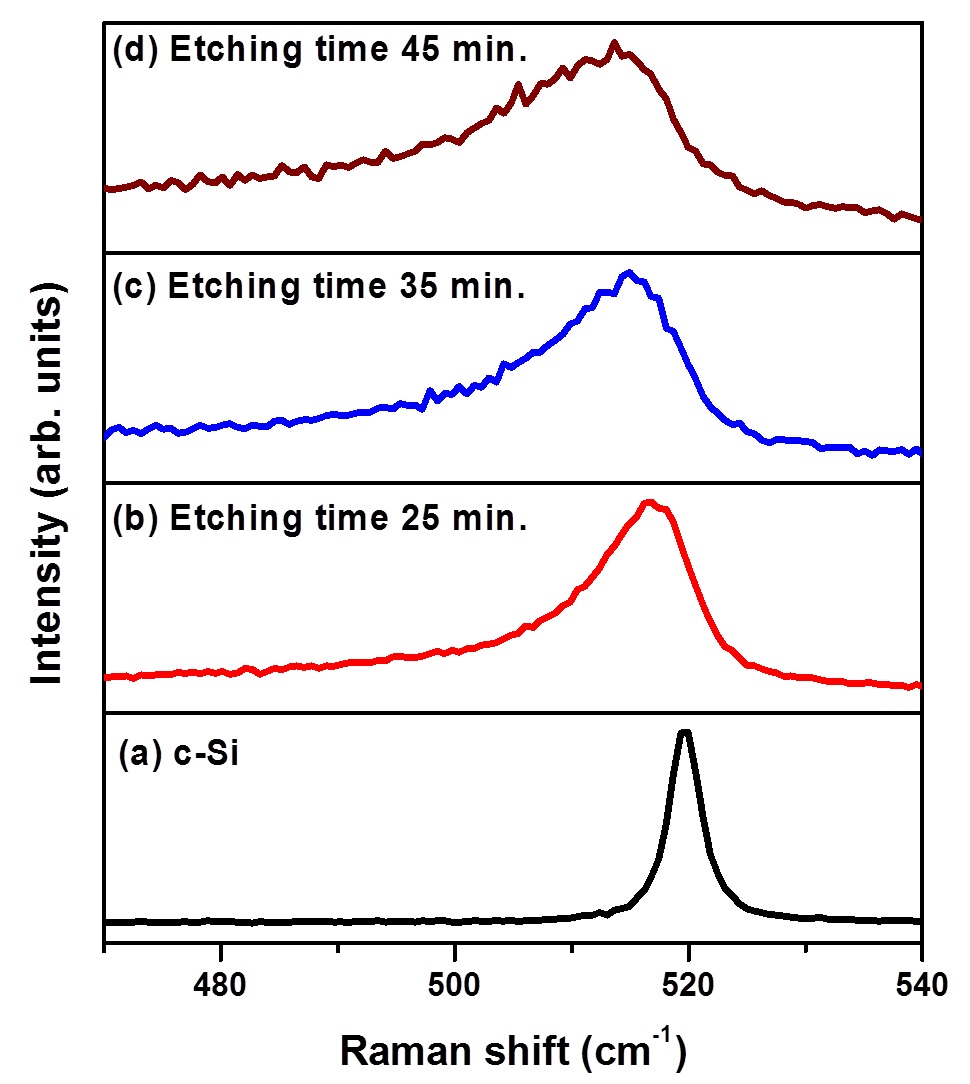}
\caption{Raman Spectra from (a) crystalline Si, (b) sample S1, (etching time: 25 minutes), (c) sample S2, (etching time: 35 minutes), and (d) sample S3, (etching time: 45 minutes).}
\end{figure}

Qualitative estimation of nanostructure size can be obtained from the Raman shifts due to the confinement effect using following relationship obtained by BPM [14]:

\begin{equation}
\Delta \omega = \omega(L) - \omega_0 = -A \left(\frac{a}{L}\right)^\gamma
\end{equation}
where $\omega(L)$ is the frequency of the Raman phonon in a nanocrystal with size $L$, $\omega_0$ is the frequency of the optical phonon corresponding to the zone-center, and a is the lattice constant of Si. The parameters $A$ and $\gamma$ are used to describe the vibrational confinement due to the finite size in a nanocrystal. The values of these two parameters, $A$ and $\gamma$ deduced by Jian Zi et.al. are 20.92 \cm and 1.08 respectively. The Raman shifts of SiNWs with respect to the frequency of the optical phonons at the zone-center versus size using Eq. (1) are shown in Fig. 3(a) by solid line. The sizes of SiNWs calculated using Eq. (1) are 3.7 nm, 2.2 nm and 1.8 nm for samples S1, S2 and S3 respectively as shown by solid circles in Fig. 3(a). Figures 3(b) \& 3(c) show the variation in asymmetry ratio and FWHM respectively with the calculated sizes of SiNWs. The asymmetry ratio and FWHM increases with decreasing size of NWs and this behaviour is same as expected from the confinement of optical phonons discussed above. The sizes of nanowires estimated by Raman scattering is not consistent with the SEM image shown in fig 1 and can be explained as follows. The broad wires ($\sim$ 100 nm) observed in Fig 1 could be bunch of wires rather than a single nanowire. This bunch in fact contains SiNWs of few nanometer size as predeicted by Raman calculations which could not be resolved by SEM.

\begin{figure}[ht]\centering 
\includegraphics[width=8cm]{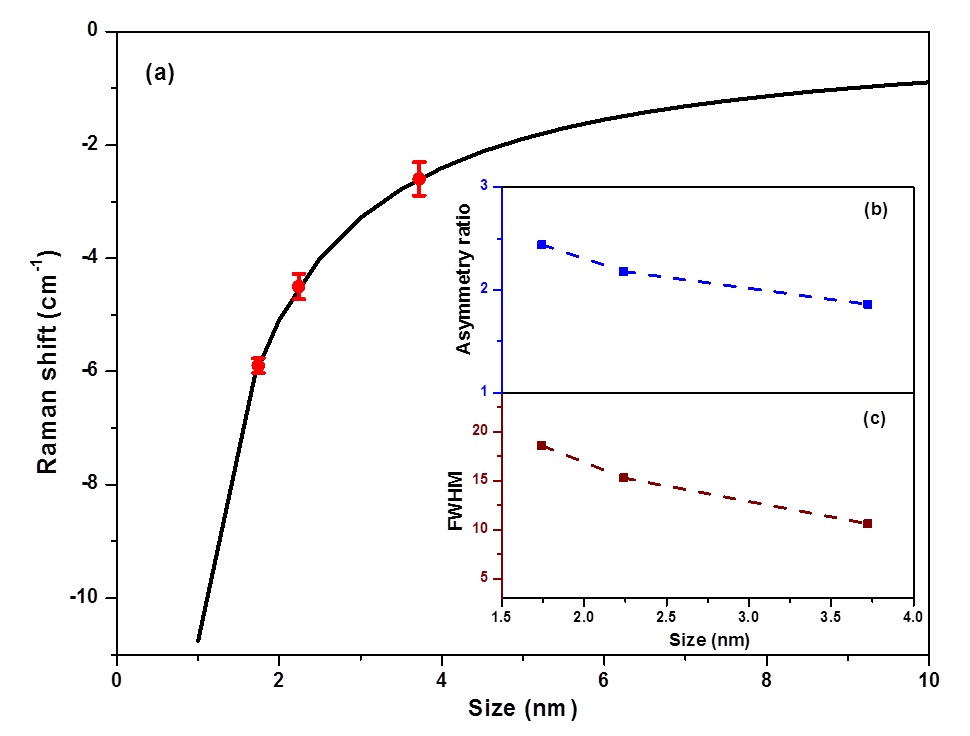}
\caption{(a) Raman shifts vs size for SiNWs. Solid line is represents the confinement model as given by Eq. (1) and solid circles are calculated results for NWs. (b) Variation in asymmetry ratio with size of NWs, (c) variation in FWHM with size of NWs.}
\end{figure}

Well aligned SiNws may have very good field emission properties and may be investigated for application in devices. In order to understand the electron field emission characteristics of vertically aligned SiNWs fabricated by MACE, the electron field emission measurements were carried out in a diode assembly kept in a vacuum chamber at a base pressure of $3\times 10^{-6}$ torr and gap between electrodes was 250 $\mu$m. The phenomenon of electron field emission is based on the effect of quantum tunneling of electrons found inside a ground conductor through a barrier that is formed by the ionic lattice of the conductor and the external electric field. Figure 4(i) shows the macroscopic current density (J) versus macroscopic electric field (E) plots for samples S1, S2 and S3. The turn-on ﬁeld for a collection of SiNWs has typically been deﬁned as the ﬁeld required for generating emission current densities of 10 $\mu A/cm^2$ and have been calculated from Fig. 4(i). The turn on field values are found to be 8.9 V/$\mu$m and 7.8 V/$\mu$m for samples S2 and S3 respectively, which is lower than other reported results34 and no turn on field is obtained for sample S1. A lowered value of turn on field necessary for field emission from samples prepared by MACE could be used for low power operation applications. In addition, MACE technique will provide a simpler technique to prepare well align SiNWs for application purposes.

\begin{figure}[ht]\centering 
\includegraphics[width=8cm]{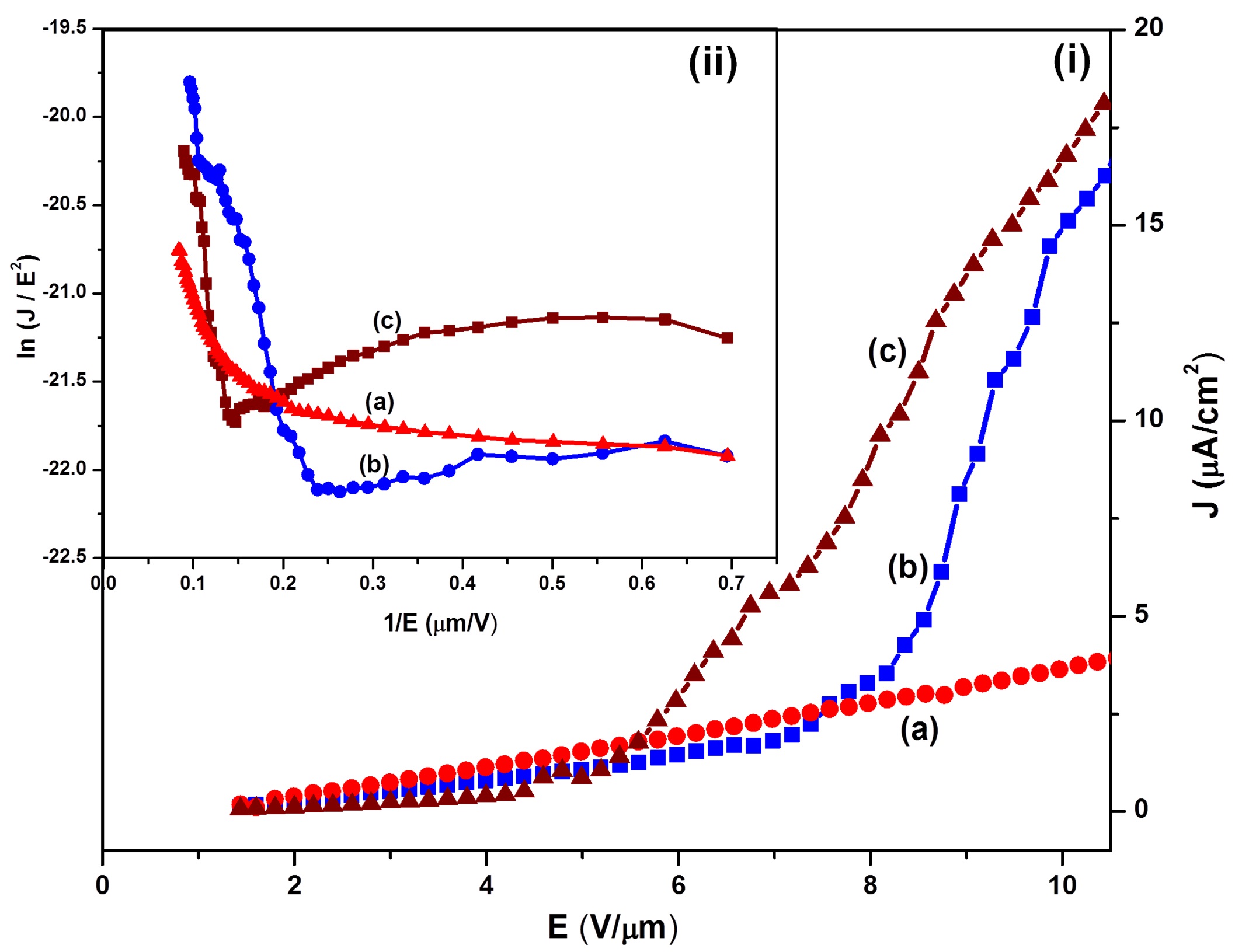}
\caption{(i) Field emission current-voltage curves from SiNWs samples S1, S2 and S3 are shown in (a), (b) and (c) respectively at measurement distance of 250 $\mu m$. (ii) corresponding  Fowler–Nordheim plots of SiNWs samples S1, S2 and S3 are shown in (a), (b) and (c) respectively. }
\end{figure}

\begin{table*}[hbt]
\caption{Field emission characteristics of SiNWs samples S1, S2 and S3 calculated by field emission data of these samples shown in Fig. 4. }
\centering
\begin{tabular}{cccc}
\hline
Sample & Size(nm) & Turn on field (V/$\mu$m) & Field amplification ($\beta$) \\
\hline
S1 & 3.7 & -- & 1140 \\
S2 & 2.2 & 8.9 & 2508 \\
S3 & 1.8 & 7.8 & 4223\\
\hline
\end{tabular}
\end{table*}

It is clear that field emission properties are different for different samples containing SiNWs of different sizes. It is easier to turn on the field emission for SiNWs having smaller diameters due to higher effective electric field at the tip of the nanowire. A simple quantum mechanical approach results in the following dependence of the emission current density $J$ on the electric field strength E, called Fowler-Nordheim (F-N) equation [35,36]

\begin{equation}
J = \frac{A \beta^2 E^2}{\phi} e^{-\frac{B \phi^{1.5}}{\beta E}}
\end{equation}

where, constants  $A = 1.54 \times 10^{-6} A eV V^{-2}$ and $B = 6.83 \times 10^3 eV^{-3/2} V/\mu m$ are known as first and second F-N constants, $\phi$ is the emitter’s local work function, $\beta$ the field enhancement factor and E is the applied electric field. The F-N plot is obtained by plotting $ln(J/ E^2)$ vs $1/ E$ as shown in Fig 4(ii). Table 1 lists the different field emission parameters of SiNWs, samples S1, S2 and S3 calculated by field emission data of these samples shown in Fig. 4. The field enhancement factor which is deﬁned as the ratio of the local electric ﬁeld at the tip of a nanowire to the macroscopic electric ﬁeld, are determined from the slope of the F-N plot Assuming that $\phi$ equals 4.15 eV for Si [37] The macroscopic field enhancement factor for sample S1, S2 and S3 are 1140, 2508 and 4223 respectively calculated from the slope of the F-N plots. Field emission characteristics show the increase in $\beta$ and decrease in turn on field with decreasing size.

All samples shows nonlinearity in the F-N plots in Fig.4 which may be due to following reasons.  (i) F-N equation serves as the foundation of cold field emission theory, it may not be suitable for predicting the emitted current density from emitters with a quantum-confined electron supply. (ii) Strictly, the F-N theory is valid for a single, metallic, planar emitter. For semiconductors, the effect of band structure, surface states and field penetration have to be accounted for in F-N theory. So some extension/ modifications are needed to explain this nonlinearity in F-N plots due to confinement and semiconductors. This study can also be extended for Ag-coated SiNWs and template based SiNWs synthesised by MACE with controlled diameter, shape, length and packing density for different potential applications in cold cathode emitters. If aligned Si nanowires with a controllable site density are obtained on ﬂat Si surface, excellent ﬁeld emission ﬂat panel display devices could be fabricated. 

\section{Conclusions}
In summary, MACE provides a simple approach to fabricate SiNWs which shows field emission at lower fields as compared to other methods to prepare SiNWs. Vertical arrays of NWs are formed across the entire surface of the wafer prepared using MACE, as shown in the cross-sectional SEM image. SiNWs prepared by this technique show quantum  confinement effects As evident from the downshifting, broadening and asymmetry of the first-order Raman line-shape. These SiNWs proved to be good field emitters and enhancement in field emission was observed with decreasing size. A size dependent change in turn on voltage indicated that field emission properties can also be tailored as a result of quantum confinement in Sinanostructures. A nonlinear F-N plot obtained from SiNWs suggests modification in this model corresponding to semiconductor, taking into the confinement effect. Our results indicate that electrochemical etching technology can provide a cheap and easier method for fabricating vertically aligned silicon nanowire arrays for potential applications in field emission devices.

\section*{Acknowledgments} 
Authors are thankful to Sophisticated Instrument centre (SIC), IIT Indore for providing SEM measurement facilities

\end{document}